\documentclass[letterpaper,12pt]{article}

\usepackage{bm}
\usepackage{color}
\usepackage{graphicx}
\usepackage{subfigure}

\begin{document}

\title{Bijels Containing Magnetic Particles: A Simulation Study}

\author{
Eunhye Kim$^1$, 
Kevin Stratford$^2$, and 
Michael E. Cates$^1$ \\
\vspace{1ex} \\
\small
$^1${\it SUPA, School of Physics and Astronomy, University of Edinburgh,}\\
\small
{\it JCMB, Kings Buildings, Mayfield Road, Edinburgh EH9 3JZ, UK}\\
\small
$^2${\it EPCC, University of Edinburgh,}\\
\small
{\it JCMB, Kings Buildings, Mayfield Road, Edinburgh EH9 3JZ, UK}\\}

\maketitle

\begin{abstract}  
Bicontinuous, interfacially jammed emulsion gels (bijels) represent a class of soft solid materials in which interpenetrating domains of two immiscible fluids are stabilized by an interfacial colloidal monolayer. Such structures can be formed by arrested spinodal decomposition from an initially single-phase colloidal suspension. Here we explore by lattice Boltzmann simulation the possible effects of using magnetic colloids in bijels. This may allow additional control over the structure, during or after formation, by application of a magnetic field or field gradient. These effects are modest for typical parameters based on the magnetic nanoparticles  used in conventional ferrofluids, although significantly larger particles might be appropriate here.  Field gradient effects, which are cumulative across a sample, could then allow a route for controlled breakdown of bijels as they do for particle-stabilized droplet emulsions.

\end{abstract}

\section{Introduction} \label{introduction}
The existence of particle-stabilized emulsions has been known for over a century \cite{ramsden}. Recently, understanding of their formation has advanced considerably \cite{binksreview}, and stimulated a wider interest in using liquid-liquid interfaces to direct the self-assembly of nanoparticles \cite{russellreview,balazs,russellpapers}. Recent simulation work in our group \cite{Stratford:2005/a,kim:2008/a} has addressed a new route to directed self assembly of colloids (including nanocolloids), creating fluid-bicontinuous structures. If the interfacial particles solidify by jamming in such a case, it should impart macroscopic rigidity in three dimensions \cite{Stratford:2005/a,japan}. The resulting `bijel' (bicontinuous interfacially jammed emulsion gel), comprising a solid matrix permeated by a pair of bicontinuous fluids, could have potential applications, for instance as a `membrane contactor' for catalytic applications \cite{Stratford:2005/a,membranecontactors,patent}.

The basic protocol for making bijels starts with a sample in the single phase region of a binary fluid mixture, in which are suspended colloidal particles having roughly equal affinity for the two fluids. There follows a quench in which the fluids demix, sweeping up particles to the interface, where they adopt a contact angle close to $90^\circ$ (neutral wetting). As the interface reduces its area by the usual coarsening process, the colloids become jammed into close proximity. There is then an energy barrier $\alpha \epsilon$, where $\epsilon = \sigma \pi a^2$, to detachment of a particle of radius $a$ from a fluid-fluid interface of tension $\sigma$, where $\alpha$ is a geometrical parameter dependent on the local particle packing. (Naively one might expect $\alpha \simeq 1$, although we have presented numerical evidence for smaller values \cite{kim:2008/a}). But unless $\alpha$ is very small indeed, the detachment barrier remains large compared to $k_BT$ for colloids larger than a few nm \cite{binksreview}, in which case the jammed bijel structure should be permanently arrested. Such fully arrested bijels have recently been created experimentally \cite{natmat,fluffybijel,sanz}, and do exhibit a yield stress. In practice the colloidal particles used -- although repulsive at moderate separations --  might be pushed by the interfacial forces into a primary flocculation minimum creating quasi-permanent bonds \cite{sanz}. However, this takes several hours (at least in the systems studied in \cite{sanz}) and so can be viewed as a final lock-in of a structure whose formation is governed by the interplay of interfacial and repulsive colloidal forces.

In this paper we present new simulations addressing the case where the colloidal particles have magnetic (or dipolar) interactions as well as the short-range repulsion considered previously. Magnetic colloids are of course well known experimentally as the basis of ferrofluids -- liquids whose material properties, particularly viscosity, are strongly sensitive to external magnetic fields \cite{rosensweig:1985/a}. A very strong field-dependence has also been demonstrated for emulsion droplets stabilized by magnetic particles \cite{fuller:2005/a}: specifically, it was found that for large (millimetre) emulsion droplets, the field gradient of an ordinary bar magnet was enough to pull magnetic particles off the fluid-fluid interface, breaking the emulsion. That work involved superparamagnetic particles rather than the permanent dipoles considered in our simulations, but the fields were high enough to give near-saturation of the dipole moments \cite{fuller:2005/a}, and in this limit the two systems should behave similarly. This suggests a wider, and so far unexploited, avenue for using magnetic particles at fluid-fluid interfaces to create field-sensitive, self-assembled materials. We explore this here in the bijel context. 

The rest of this paper is organized as follows. In Section \ref{methods} we give a brief account of the simulation methodology; this combines elements from our earlier work on bijels \cite{kim:2008/a} and on magnetic colloids \cite{kim:2009/a}.  We then quantify in Section \ref{morph} the interfacial morphologies of bijels and droplets containing magnetic particles, created in the presence or absence of a uniform magnetic field. In Section \ref{deform} we consider cases where a field or field gradient is applied late in the formation of a bijel, and investigate the extent of the resulting structural perturbations. Section \ref{conclude} contains a discussion.

\section{Simulation methods} \label{methods}

We use the lattice
Boltzmann (LB) method\cite{succi_ref17} for a binary fluid\cite{swift_ref18} incorporating spherical solid particles 
\cite{jsp}, modifying a standard
`bounce-back on links' method \cite{ladd_ref19,nguyen}, to allow
for the presence of a binary solvent \cite{jsp,desplat}. 
The two
solid-fluid interfacial tensions are exactly equal, with the interfacial thermodynamics implemented as reported previously \cite{jsp}.
We choose for simplicity a pair of fluids with equal density $\rho$ and viscosity $\eta$.
Their phase diagram is also symmetric, and
described by the free energy functional \cite{kendon}
\begin{equation}
F[\psi] = \int dV \left(A\psi^2/2 + B\psi^4/4 + \kappa (\nabla\psi)^2/2\right) \label{freeE}
\end{equation}
where the order parameter $\psi$ describes the fluid composition, and
the model parameters $A<0$, $B$, and $\kappa$ control the
fluid-fluid interfacial tension $\sigma$ and thickness $\xi$ 
\cite{kendon}. The $\psi = 0$ isosurface defines the position of the fluid-fluid interface.

The binary fluid is initialized to be well mixed and at rest, with a small amplitude
random noise added to the $\psi$ field to
induce spinodal decomposition. The colloids are initially
positioned at rest randomly throughout the system: in effect therefore we quench from infinite temperature.
We choose a deep quench in the sense that thermal fluctuations of $\psi$ remain negligible: without particles, the phase separation proceeds purely by amplification of the initial noise. Time-dependent, thermal noise is however fully
included in the description of fluid momentum, using a method reported previously \cite{ronojoy}. As a result, the colloids undergo realistic, many-body Brownian motion. 

Unless otherwise stated below, our parameters are set as follows. First, we choose $-A=B=0.002$, and $\kappa = 0.0014$, giving an
interfacial thickness of $\xi = 1.14$ and tension
$\sigma = 1.58\times 10^{-3}$; we set the (scalable) fluid density $\rho = 1$, and choose
viscosity $\eta= 0.1$. Here and below, all physical variables are expressed in lattice units \cite{kendon}, except where SI units are given explicitly. For numerical efficiency, the colloid radius $a$ is made as small as is compatible with acceptably accurate simulation \cite{Stratford:2005/a,codef,nguyen}; we choose $a=2.3$, so that $\epsilon = 0.026$. Our default noise setting is $k_BT = 2.13\times 10^{-5}$, so that $\epsilon/k_BT = 1230$. 
We can choose to interpret these parameters as representing a short-chain hydrocarbon/water or hydrocarbon/alcohol mixture, with laboratory parameters 
$\rho = 1000$~kg~m$^{-3}$, $\eta = 9.3\times 10^{-4}$~N~m$^{-2}$s 
(close to that of water at $\sim 300$K) and $\sigma = 6.1\times 10^{-2}$N~m$^{-1}$. This corresponds to a physical particle radius of $a=5.1$ nm. 
We have in addition performed several runs at a temperature $k_BT=2\times10^{-4}$, approximately ten times larger than the default value. This allows us effectively to probe longer timescales (since the Brownian diffusivity increases by the same factor) while maintaining a value of the capillary energy that is still very large compared to thermal energies ($\epsilon/k_BT\simeq 130$).

The volume fraction of colloids is $\phi=0.20$ in all simulations reported here. However, in the bijel these particles get packed into close proximity forming a network of repulsive contacts. LB does not properly resolve the lubrication films at these contacts; to correct for this we introduce a normal lubrication force for surface separations $h = r-2a<h_c = 0.3$. Such lubrication forces slow the code down considerably once there are large clusters of particles in mutual lubrication contact; to minimize this, a soft-core thermodynamic repulsion is introduced so that interfacial particles maintain a separation somewhat larger than $2a$ \cite{Stratford:2005/a}. (Such a repulsion is often present physically, e.g., for charge-stabilized colloids.) 
The chosen `soft-core' short-range potential $U^{sc}(h)$ as a function of the particle surface separation $h = r-2a$ obeys 
\begin{equation}
U^{sc}(h) = u(h) - u(h_c) -(h-h_c) \left(\frac{du}{dh}\right)_{h=h_c}
\end{equation}
where $u(h) = \gamma(h_0/h)^\nu$, and $h_c$ is a cut-off separation. (Thus $U^{sc}$ is a `cut-and-shift' power law potential which vanishes smoothly at $h_c$.) The parameter set is chosen as $\gamma=10k_BT$, $h_0=0.1$, $\nu=1.0$ and $h_c=0.25$, as benchmarked previously \cite{kim:2008/a,kim:2009/a}. 
All simulations have been done on a $D3Q19$ lattice with the system volume $\Lambda^3=64^3$ and with either periodic or fixed-wall boundary conditions. For the latter conditions, a normal lubrication force is introduced for colloids at distances $h < h_{lub}=0.5$ between the plane of the wall and the surface of the colloid; a neutral wetting condition is maintained between such boundary walls and the fluid-fluid interfaces.

A dipole in an external magnetic field ${\bf B}$ is subject to a force ${\bf F}_i = m_i \hat{\bf s}_i.\nabla {\bf B}$ and a torque ${\bf T}_i = m_i \hat{\bf s}_i\times {\bf B}$, where $m_i$ is the magnitude of dipole moment $i$ and $\hat{\bf s}_i$ its direction. (We take $m_i = m$ for all particles.)
The dipole-dipole interaction is given directly as
\begin{equation} \label{dipole}
U_{ij}^{d} ({\bf r}_{ij}) =  \frac{\mu_0}{4\pi}\frac{m_im_j}{r_{ij}^3} \left[ {\hat {\bf s}}_i \cdot {\hat {\bf s}}_j - 3({\hat {\bf s}}_i \cdot \hat{{\bf r}}_{ij} )( {\hat {\bf s}}_j \cdot \hat{\bf r}_{ij} ) \right]
\end{equation}
where the prefactor $\mu_0/4\pi = 10^{-7}$H/m.
The relative importance of the dipole-dipole interaction, compared to thermal energies, is characterized by the dimensionless parameter
\begin{equation}
\lambda = \frac{\mu_0}{4\pi} \frac{m^2}{8a^3k_BT}.
\end{equation}
For ferrofluid applications, $\lambda$ values in the range $2-8$ are readily available but larger values remain relatively uncommon (see \cite{kim:2009/a} and references therein).
To calculate the long-range dipolar interactions for a periodic simulation box, Ewald summation is used following the procedure and parameter choices laid out in \cite{kim:2009/a}: this is unaffected by the switch from a single solvent in that paper to a binary one here. In the case of fixed wall boundary conditions the long-range dipolar interaction was directly calculated from Eq.\ref{dipole}. 

Below we examine the behavior of bijels stabilised by magnetic colloids under either a uniform magnetic field or a field gradient. A uniform field ${\bf B}_0$ directly creates torque but not net forces on the dipoles; by aligning the dipoles, it alters the dipole-dipole interactions. A uniform field applied within the simulation box clearly obeys ${\bf \nabla}.{\bf B}_0=0$, and can be used with either periodic or fixed-wall boundary conditions. In the case of a field gradient, we restrict attention to fixed-wall boundary conditions since the requirement of  ${\bf \nabla}.{\bf B}_0=0$ is then not easily satisfied in the periodic case. The field pattern chosen is in fact that of a point magnetic dipole (much larger in magnitude than those on the particles) held at a fixed orientation at a location outside the simulation box. 

Most of our simulations have been evolved for $5\times10^{5}$ timesteps; this is on the microsecond timescale for nanocolloids \cite{Stratford:2005/a} and is long enough to reach a near-steady state with our chosen lattice size ($\Lambda^3=64^3$), at least so long as one disregards the slow aging effects exploreded in \cite{kim:2008/a}. For one such run, a single Intel 2.4GHz processor requires 270 hours. Parallel computation using 8 cores of a cluster of 3GHz Intel Dual-core processors reduce that computational time to about 26 hours. 

\section{Fluid and magnetic morphologies}

\label{morph}
In this Section we present the morphologies of fluid domains arising under various conditions. We vary the initial volume ratio of fluids ($\psi_0$), the dipolar coupling constant ($\lambda$), and the strength of a uniform external field (${\bf B}_0$). We always use the same initial conditions with a uniformly mixed fluid phase and random positions of colloids. We use the default temperature value except where otherwise stated.  

\subsection{Bicontinuous phase and droplet phase}

The fluid domain morphologies resulting from our chosen quench protocol have been presented elsewhere for nonmagnetic colloids ($\lambda =0$) \cite{kim:2008/a}. By varying the composition variable $\psi_0$ away from the symmetric value $\psi_0=0$, the percolation threshold $\psi_p$ between bicontinuous and droplet morphologies was estimated to lie within the range from $0.3$ to $0.4$. 

\begin{figure}[tbph]
 \centering
\subfigure[]{\label{fig1a}\includegraphics[width=0.75\textwidth]{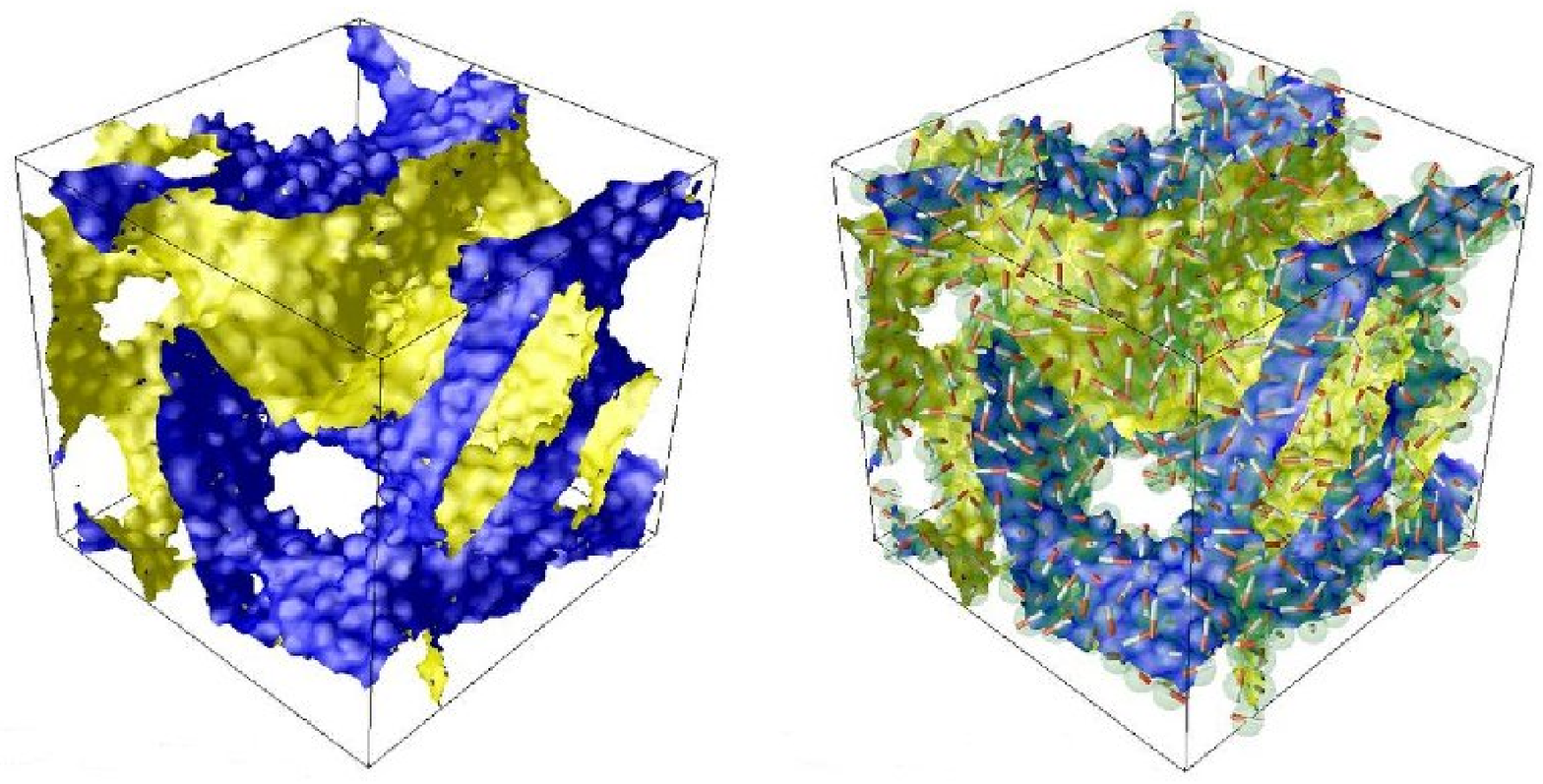}}
\subfigure[]{\label{fig1b}\includegraphics[width=0.75\textwidth]{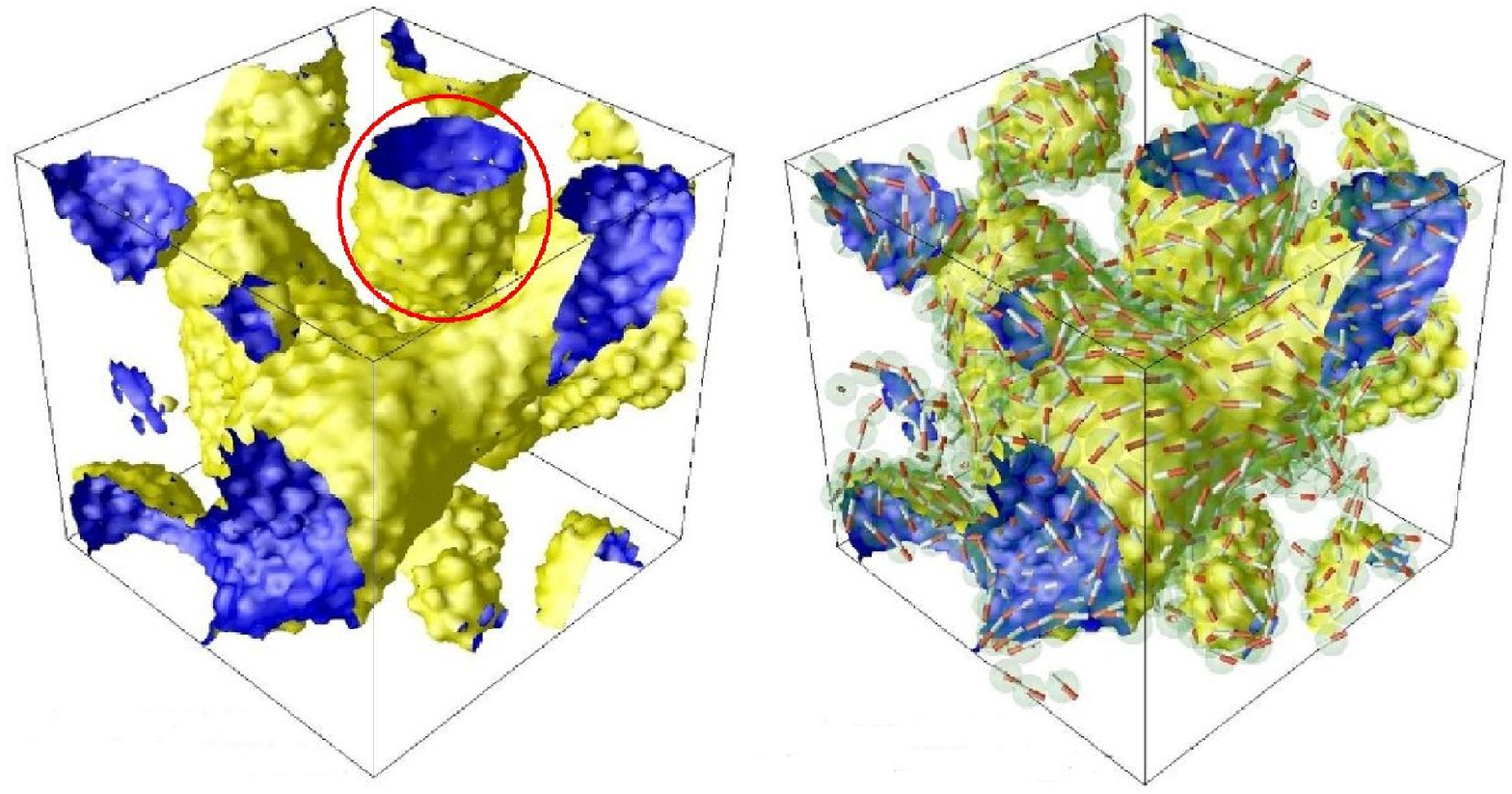}}
\caption{The morphologies in a system with periodic boundary conditions for $\lambda=4$ at $k_BT=2.133\times10^{-5}$; system size is $\Lambda^3=64^3$. The right panels shows the fluid-fluid interfaces plus colloids represented as translucent spheres of radius $a$. Within each such sphere, a rod is shown to represent the magnetic dipole whose poles are painted red and white. The left panels show the same images without the particles to enable fluid domains to be seen more clearly. (a) Symmetric quench: $\psi_0=0.0$. (b) Asymmetric quench: $\psi_0=0.4$. One droplet (circled in red) is chosen to see the geometry of magnetic particles on the surface, discussed in Section \ref{droplet}.}
\label{fig1}
 \end{figure}

Figure \ref{fig1} presents snapshots for the domain structure at the end of the simulation with a dipolar coupling constant $\lambda=4$ for both a symmetric and a strongly asymmetric ($\psi_0=0.4>\psi_p$) quench, corresponding to fluid ratios of 50:50 and 70:30 respectively. (We use periodic boundary conditions; results for closed-box boundary conditions are similar.) In each case the basic morphology is broadly similar to that found for $\lambda =0$ in previous work \cite{kim:2008/a}. This is not surprising since, although the magnetic interaction is several times larger than $k_BT$, the balance between capillary forces and hard-core repulsions involves a much higher energy scale than this ($\epsilon = 1230k_BT$). Within the particle layer, however, the magnetic interactions promote the dipolar particles to align in a nose-to-tail fashion in all cases; this requires the dipole vectors to lie tangent to the local monolayer. The curvature of this monolayer then prevents the formation of a uniform monodomain of particles, creating frustration. Thus in the absence of an external field we see no long range order and expect zero mean magnetization (although in practice a small nonzero value is measured due to finite size effects). In Fig.\ref{fig1} we identify a representative droplet within the depercolated regime; in Section \ref{droplet} we study more carefully the local magnetic ordering at the surface of this droplet.

We have tried increasing $\lambda$ up to $\lambda = 40$, well above the usual experimental range for magnetic colloids. We find that for symmetric quenches, although the magnetic ordering within the particle layer increases (longer chains of head-to-tail contacts) the bicontinuous morphology of fluid domains is not significantly altered. The same applies for the asymmetric quench at $\psi_0 = 0.3$ whose structure can be seen in Fig.\ref{fig6}(a) below, where it represents an initial state prior to switching on a magnetic field.

\subsection{Droplet covered by magnetic particles}
\label{droplet}

\begin{figure}[tbph]
\centering
\subfigure[]{\label{fig2a}\includegraphics[width=0.7\textwidth]{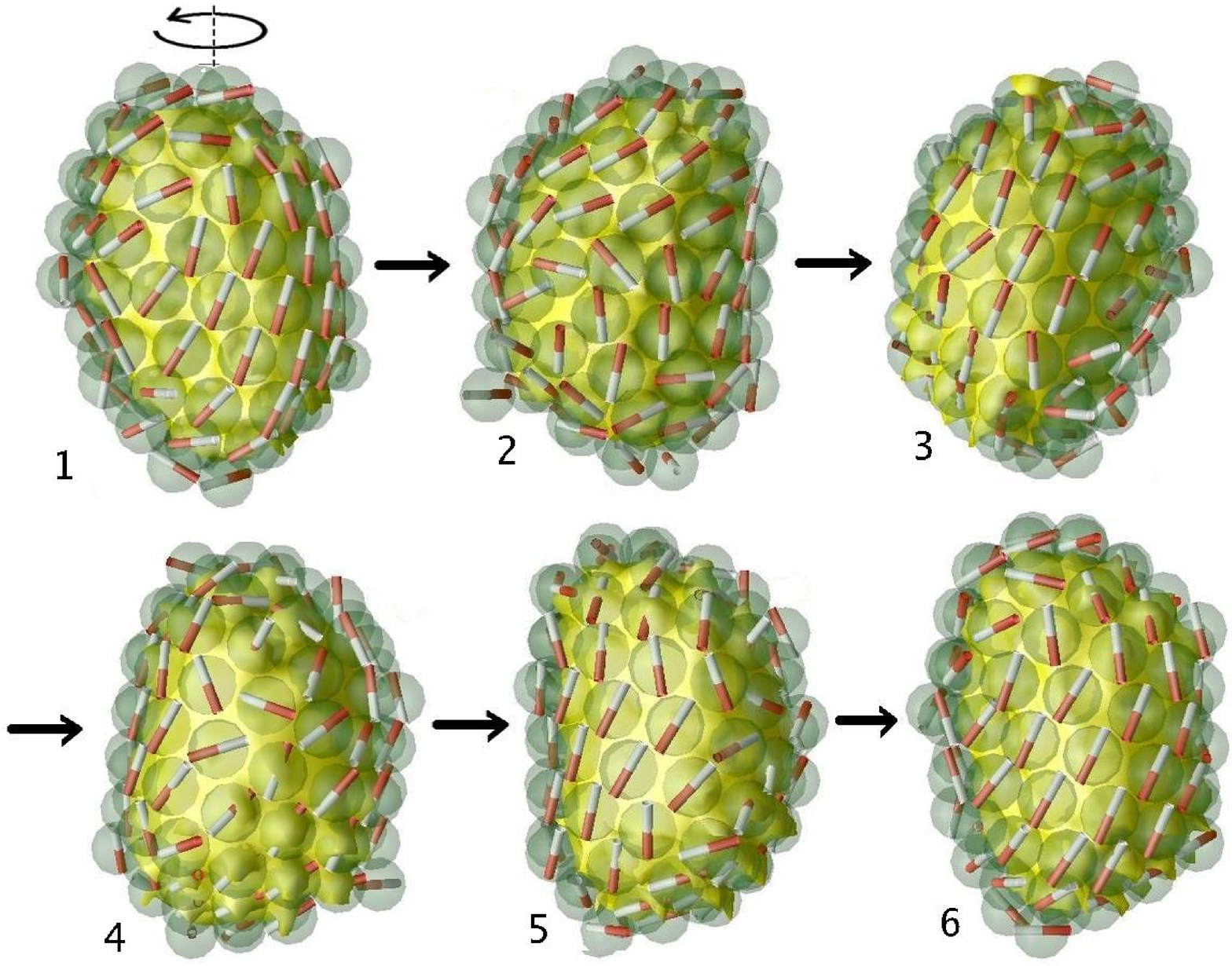}}
\subfigure[]{\label{fig2b}\includegraphics[width=0.2\textwidth]{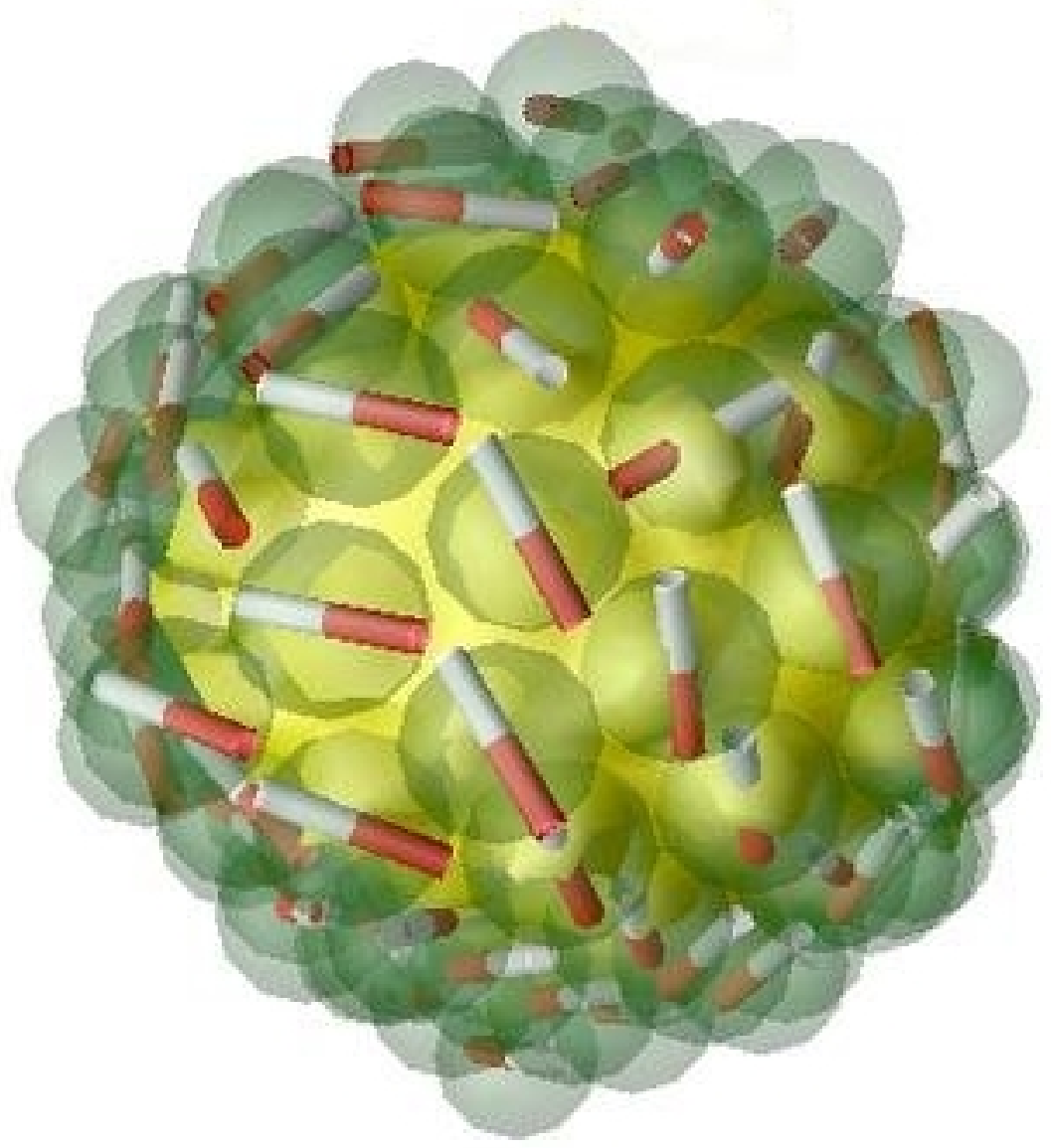}}
\subfigure[]{\label{fig2c}\includegraphics[width=0.4\textwidth]{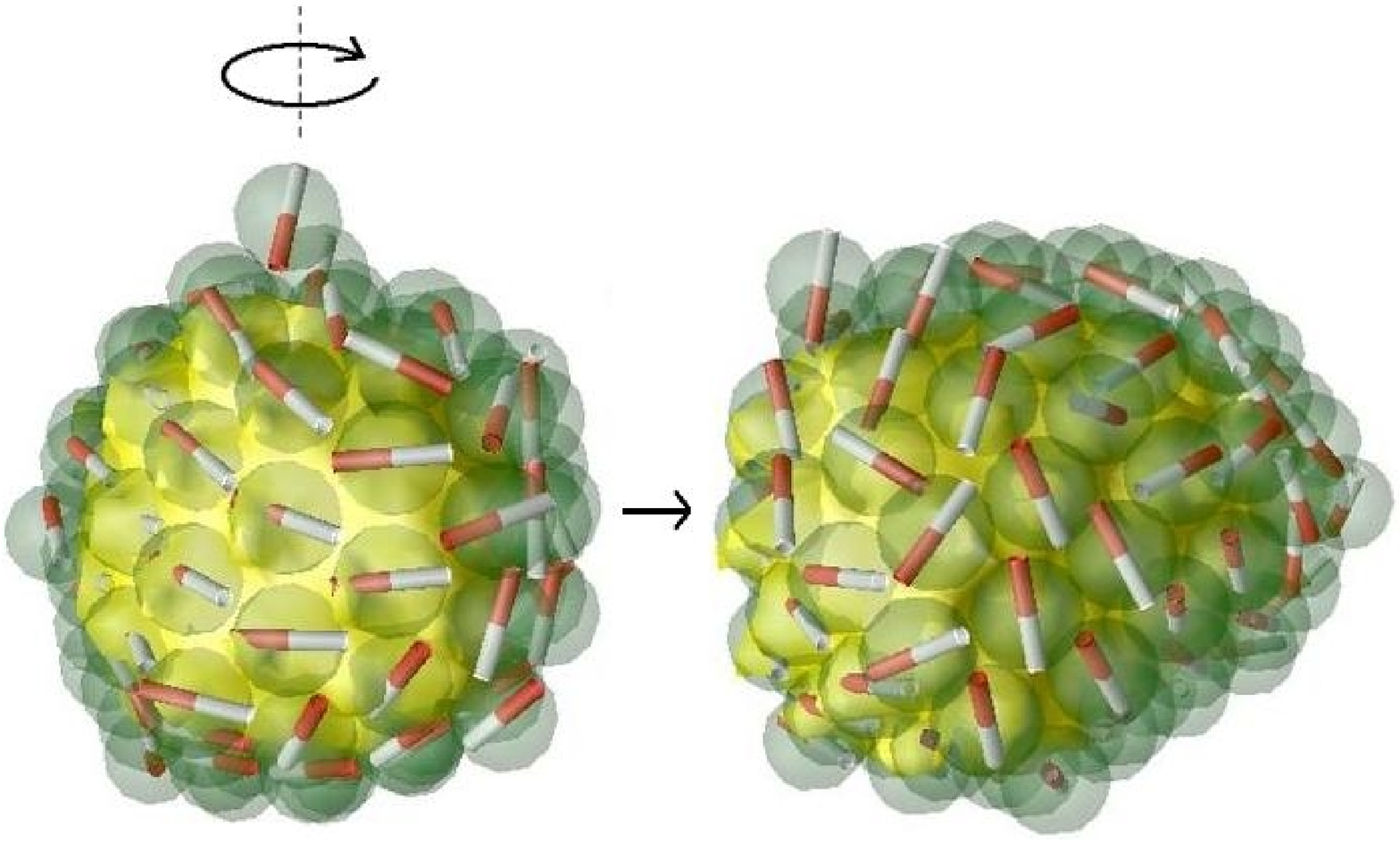}}
\subfigure[]{\label{fig2d}\includegraphics[width=0.25\textwidth]{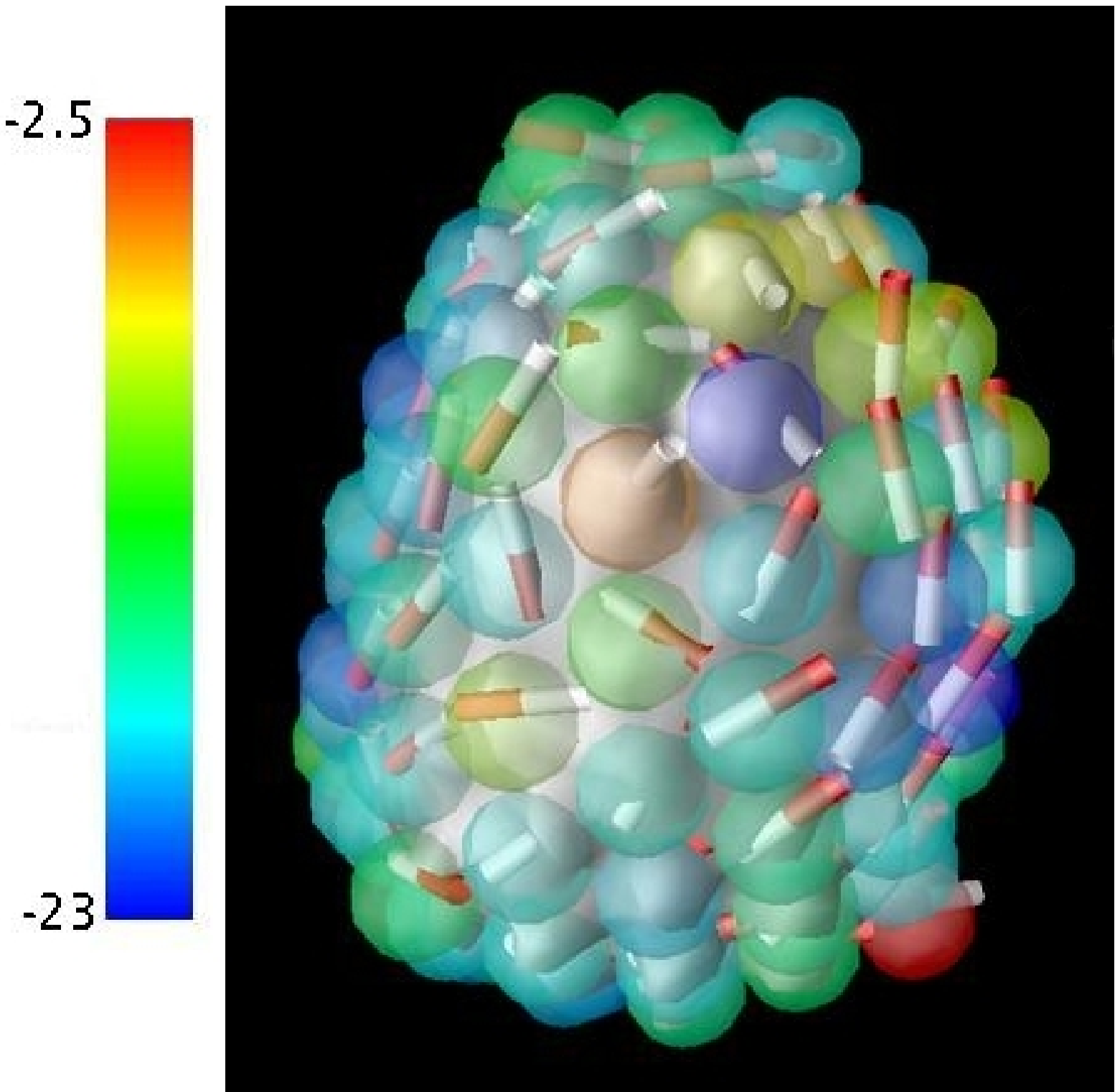}}
\subfigure[]{\label{fig2e}\includegraphics[width=0.3\textwidth]{fig2e.eps}}
\caption{Dipolar ordering on the surface of the droplet. (a) Snapshots for sides of the droplet rotating counter-clockwise. (b) Viewed from above. (c) Viewed from below.
(d) The directions of dipoles covering the droplet are shown with particles coloured according to $U^d_i/k_BT$. A frustrated particle (orange) is found among dipoles on the surface, having dipolar energy $U^d/k_BT=-4.70$. (e) The local dipolar energy distribution of surface colloids.}
\label{fig2}
 \end{figure}

Figure \ref{fig2} shows a droplet chosen from the snapshot in Fig.\ref{fig1}(b), where the magnetic particles on its surface have aligned to reduce their interaction energies. As discussed in ref. \cite{kim:2008/a}, this droplet is aspherical. Fig.\ref{fig2}(a) shows the droplet with different viewpoints rotating by a full turn counter-clockwise; hence pictures 1 and 6 are the same. The particles achieve a local packing that is mostly close to hexagonal ordering, with dipolar orientations primarily nose-to-tail, following the curvature of the surface. In pictures 1 and 3, domains of parallel magnetic alignment are seen, but with opposite orientations on the two faces of the droplet. 
The intermediate viewpoints (2 and 4) show nontrivial defect structures, as do 
Fig.\ref{fig2}(b,c) showing the top and bottom of the droplet. This arrangement of dipoles shows considerably more frustation than strictly necessary to accommodate a locally aligned vector field on a surface of spherical topology (see, e.g. ref. \cite{park_ref27}). Although the observed magnetic alignment might anneal somewhat if the simulation were run for much longer, at least some of these defect structures are likely to be permanently trapped as a result of the underlying defects in the particle packing. Note that for a perfect hexagonal lattice, there is a magnetic energy gain of $4 \lambda k_BT$ per particle in creating a perfect in-plane head-to-tail alignment compared to one in which the dipoles point at 30 degrees (but still in plane) to the lattice vector; but with a defect-ridden packing (locked in by capillary forces) local as well as global frustration may prevent this energy gain from being realized. 

In general, the dipolar energy for an individual particle can be defined from Eq.\ref{dipole} as $U^d_i = \sum_jU^d_{ij}$, where to maintain consistency with periodic boundary conditions, the sum is restricted to  $r_{ij}<\Lambda/2$. In the Ewald method the remaining contribution is computed in Fourier space, but in the absence of long range order this term should not dominate, and we can usefully study the distribution of local energies with it neglected. We show in Fig.\ref{fig2}(d) an image of the droplet with particles color-coded according to their magnetic energy, and in Fig.\ref{fig2}(e) a histogram of the dipolar energies observed. The single most frustrated particle (full red, bottom right in (d)) is in fact not attached (or barely so) to the fluid-fluid interface. (The method used to identify particles includes all those whose centres lie within a certain distance of the interface, but depending on the arrangement of other particles this may or may not signify attachment.) Of the fully attached particles the most frustrated is at upper left centre (translucent orange) with an energy of $-4.7k_BT$. This is surrounded by a hexagonal ring on which the dipolar axis rotates through a full turn. 

\subsection{Effect of uniform external field}

In the absence of an applied external field, magnetic dipoles have only local orientational correlations. However, under the external field, dipole moments preferentially align along the field direction. At very high field strength, completely aligned dipoles can be expected. At intermediate values, the magnetic saturation for noninteracting permanent dipoles is predicted by Langevin theory
\cite{rosensweig:1985/a} as a function of $\alpha_B= m B/k_BT$. A recent study of ferrofluids in a uniform field \cite{Ivanov:2007/a} has shown that the noninteracting Langevin theory is still applicable in systems with moderate dipole-dipole interactions.

\begin{figure}[tbph]
 \centering
\subfigure[]{\label{fig3a}\includegraphics[width=0.75\textwidth]{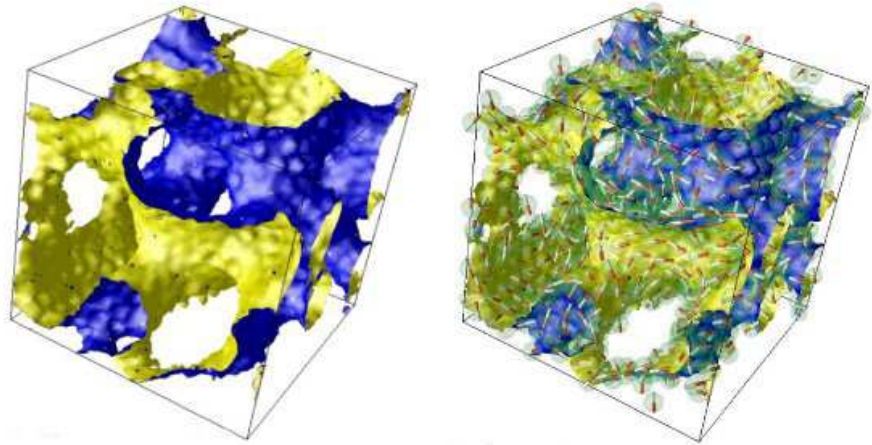}}
\subfigure[]{\label{fig3b}\includegraphics[width=0.75\textwidth]{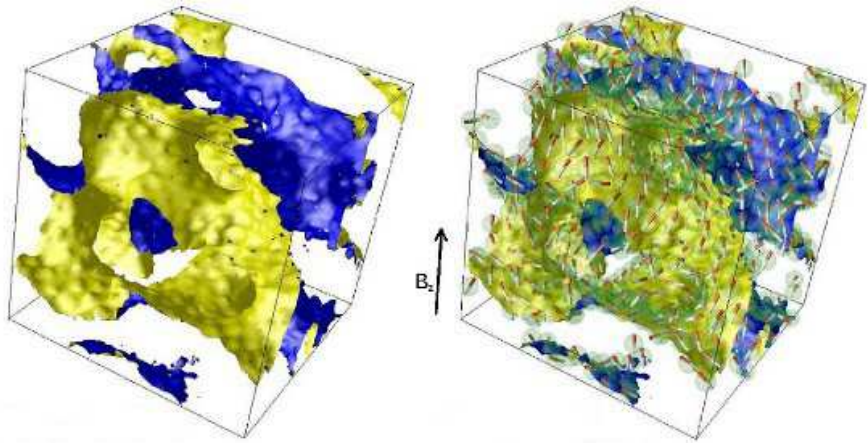}}
\subfigure[]{\label{fig3c}\includegraphics[width=0.75\textwidth]{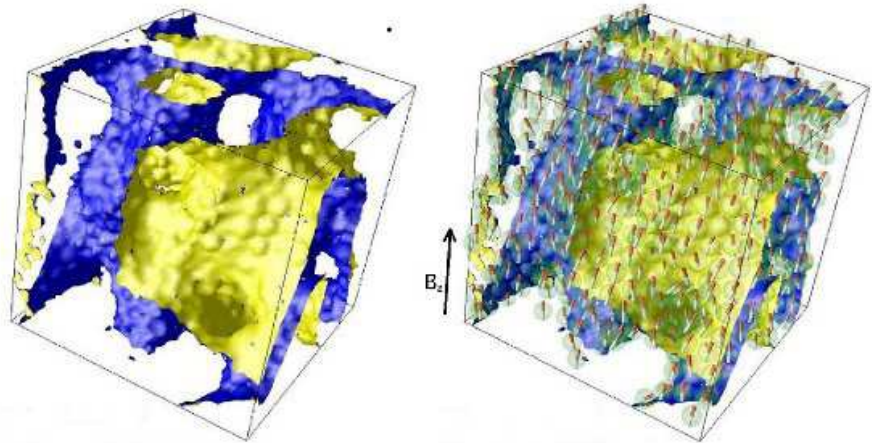}}
\caption{Final morphologies for $\lambda=4$ at $k_BT=2\times10^{-4}$. System size is $\Lambda^3=64^3$; periodic boundary conditions are used. (a) In zero field. (b) With dimensionless field strength $\alpha_B=2$. (c) With $\alpha_B=20$.}

\label{fig3}
 \end{figure}

Figure \ref{fig3} shows domain morphology snapshots for simulations in which the quench was made with a uniform magnetic field present from the outset, with $\alpha_B=2$ and $20$ respectively and with $\lambda=4$. To speed approach to the final, locally equilibrated structure, we performed these runs with enhanced diffusion (achieved by setting $k_BT = 2\times 10^{-4}$). The applied uniform field is given as ${\bf B}=(0, 0, B_z)$ and the corresponding Langevin values for the magnetic saturation $\sum_i(\hat{\bf s}_i)_z/N$ are $s=0.54$ and $s=0.95$. The measured values are instead $s=0.69$ and $s=0.93$, presumably reflecting the combined effects of surface confinement and dipolar interactions. At $\alpha_B=2$, the initial domain morphology is only weakly perturbed by the interactions of dipoles either with each other, or with the external field. At the higher field strength, $\alpha_B = 20$ corresponding to $B=0.31$T in the laboratory, significant distortion of the fluid-fluid interface is seen, stretching it along the field direction. Although the capillary and hard-core forces remain strong (we now have $\epsilon/k_BT \simeq 130$ for the capillary energy), at these field levels the requirement of dipolar alignment with the field, combined with the energetic preference for head-to-tail dipolar chains, can compete with them. However, the degree of distortion remains relatively modest, and, at least for a symmetric quench, fluid bicontinuity is maintained.

\subsection{Time evolution of characteristic domain lengths}

The morphology of the bijel structure in the presence of a field can be characterized via length scales $L_{x,y,z}$ defined \cite{Wagner:1999/a} as $\xi L_x = 1/d_{xx}$, etc., where $d_{\alpha\beta} = \sum_{\bf r} \partial^D_\alpha\psi({\bf r})\partial^D_\beta\psi({\bf r})/\sum_{\bf r} \psi^2({\bf r})$. Here $\partial^D$ is a symmetric discrete derivative and the sum is over all fluid sites. These characteristic lengths measure in essence the mean distance between interfaces along a line drawn along each of the coordinate axes; with a field along $z$, directions $x,y$ should be statistically equivalent. (However, once the domain size becomes comparable to the simulation box, their $L$ values may differ significantly in any instance.) For
the case of zero field, the structure is isotropic and a more convenient measure of the length scale is $L(t) = 2\pi\int S(k)dk/\int kS(k)dk$ where $S(k)$ is the usual density-density correlator \cite{kendon}.

\begin{figure}[tbph]
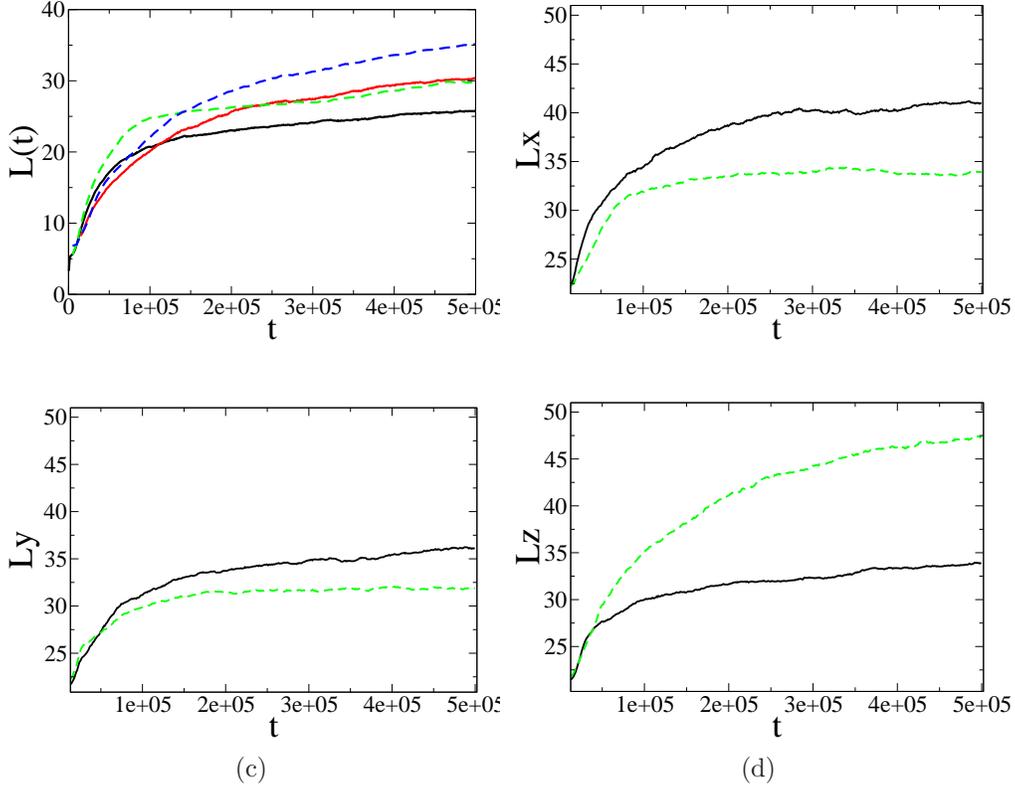

\centering
\vspace{.5in}
\subfigure[]{\label{fig4a}\includegraphics[width=0.4\textwidth]{fig4a.eps}}
\subfigure[]{\label{fig4b}\includegraphics[width=0.4\textwidth]{fig4b.eps}}
\subfigure[]{\label{fig4c}\includegraphics[width=0.4\textwidth]{fig4c.eps}}
\subfigure[]{\label{fig4d}\includegraphics[width=0.4\textwidth]{fig4d.eps}}
\caption{(a) The characteristic length scale, $L(t)$ for magnetic bijels in zero field. Black and red solid lines are $\psi_0=0.0$ and $0.4$ respectively, with $\lambda=4$. Green and blue dashed lines are $\psi_0=0.0$ and $0.4$ with $\lambda=0$. Frames (b--d) show $L_{x,y,z}$, the characteristic length scales in an applied uniform field, for dipole-dipole strength $\lambda = 4$. The black solid line has dimensionless field strength $\alpha_B = 2$; the green dashed line has $\alpha_B=20$. For the parameters selected (here with $k_BT = 2 \times 10^{-4}$, 40 lattice units corresponds to 88.4\rm{nm}, while 500,000 time steps corresponds to 275\rm{ns}, or about 5 Brownian times (the time taken for the particle to diffuse its own radius).)} 
\label{fig4}
\end{figure}

Figure \ref{fig4}(a) presents the time evolution of $L(t)$ for symmetric and asymmetric quenches in the absence of a field, with $\lambda = 4$ and, for comparison, $\lambda = 0$. At the chosen (higher) temperature, the correlated relaxation of dipolar particles leads to a decrease in the domain length scale $L(t)$ at given time, compared to the case without dipoles. (At the default temperature, not shown, there is instead a modest increase.)
Fig. \ref{fig4}(b-d) show $L_{xyz}$ for symmetric quenches in uniform fields of strength $\alpha_B = 2,20$ again with $\lambda = 4$. As visible already in Fig.\ref{fig3},  whereas for $\alpha_B = 2$ any elongation lies within the statistical scatter arising from a near-isotropic structure, for the higher field we find
significant stretching along the field direction with a domain length along $z$  about $50\%$ larger than in the transverse directions.
 
\subsection{Energetic relaxation of magnetic colloids}

\begin{figure}[tbph]
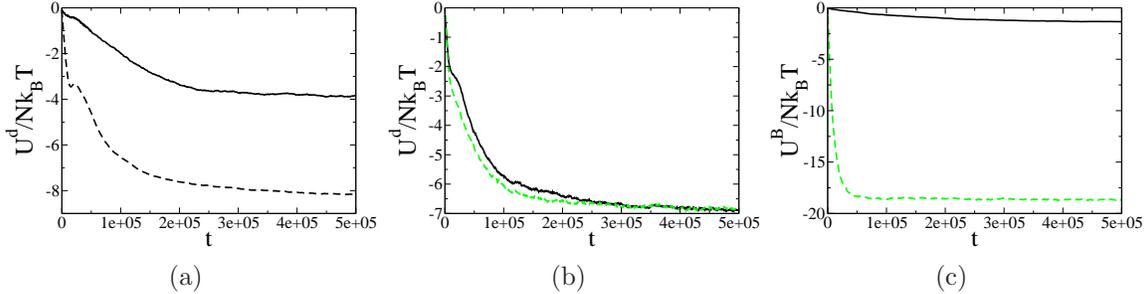

\centering
\vspace{.5in}
\subfigure[]{\label{fig5a}\includegraphics[width=0.3\textwidth]{fig5a.eps}}
\subfigure[]{\label{fig5b}\includegraphics[width=0.3\textwidth]{fig5b.eps}}
\subfigure[]{\label{fig5c}\includegraphics[width=0.3\textwidth]{fig5c.eps}}
\vspace{.05in}
\caption{ (a) The dipole-dipole energy per particle $U^d/Nk_BT$ at $\lambda=4$ for $\psi_0=0.0$ (solid line) and $\psi_0=0.4$ (dashed line) with $k_BT = 2.133\times 10^{-5}$. This shows incomplete relaxation of the magnetic degrees of freedom, as revealed by the remaining two plots for which $k_BT = 2\times 10^{-4}$, giving enhanced rotational diffusion of the particles without creating a significant tendency of colloids to leave the interface \cite{kim:2008/a}. These show (b) $U^d/Nk_BT$ and (c) $U^B/Nk_BT$ at $\lambda=4$, $\psi_0 = 0$;  the black solid line has $\alpha_B = 2$ and the green dashed line has $\alpha_B=20$. }
\label{fig5}
\end{figure}

Our bijel quenches always start with the colloids positioned at random; in the magnetic case this corresponds to a high temperature limit ($\lambda = 0$) as the initial state. Thus when the quench is performed, the fluid phase separation and the aggregation tendency of dipolar colloids are switched on concurrently. As in the nonmagnetic case, particles become jammed at the fluid-fluid interfaces causing arrest of $L(t)$ and then a slow aging of this quantity. With the temperature at its default value, we observe in addition a very slow relaxation of the dipole-dipole energy, Fig.\ref{fig5}(a), suggestive of incomplete equilibration of the magnetic degrees of freedom, even under conditions where the particle packing is considered fixed. As explained above, the magnetic relaxation can be speeded up by increasing the diffusivity using an enhanced temperature ($k_BT = 2\times 10^4$) while retaining a large value $\epsilon \sim 130k_BT$ for the capillary energy scale. Once this is done, the observed behavior is consistent with relaxation to a state of local equilibrium for the given particle packing; the dipolar energy continues to evolve slowly, but at a rate set by slow aging of the packing itself \cite{kim:2008/a}. Fig.\ref{fig5}(b,c) show the resulting relaxations of the dipolar energy and also the magnetic field energy per particle, in quenches at $\lambda = 4$ with $\alpha_B = 2$ and $20$. The field energy shows a similar aging process in a weak field but, at $\alpha_B=20$, saturates much sooner. This is not surprising as for strong fields the particles will align fully with the field direction regardless of their disposition at the interface. 
The final field energies per particle (with $U^B = -m\sum_i\hat{\bf s}_i.{\bf B}_0$) are $U^B/Nk_BT = -1.37$ and $-18.65$ for $\alpha_B = 2,20$ respectively; these values follow directly from the magnetic saturation values $s=0.69$ and $s=0.93$ quoted previously.

Table I presents the dipole-dipole and short-range repulsive contributions to the mean energy per particle for all runs so far discussed. Both quantities are averaged over a time window comprising the last $50,000$ timesteps in each run. This averaging improves the statistics, although significant run-to-run variations remain (particularly in the default temperature samples, which are not fully equilibrated magnetically). This is not surprising since the arrested domain length-scales $L_{xyz}$ are of order half the simulation box size, so the system is not self-averaging. For magnetic colloids in a single fluid \cite{kim:thesis}, the equilibrium dipolar and soft-core energies at $\lambda=4$, $\alpha_B=0$ and $\phi=0.20$ are measured as $U^d/k_BT=-4.904 \pm 0.004$ per particle and $U^{sc}/k_BT=0.0326\pm 0.0001$. Values of the latter are higher in magnetic bijels than in conventional ferrofluids, as a result of the jamming of colloids into close contact at the fluid-fluid interface. Among equilibrated samples, the dipolar energies are made somewhat more negative by the application of a field, consistent with a suggestion that the field modestly reduces the magnetic frustration of the particles at the interface.

\begin{table}[t]
{\small
\begin{tabular}{c c c c c c c c}
\hline
\hline
$\psi_0$ & $k_BT$ & $\lambda$ & $\phi$  & $\alpha_B$ & & $U^d/Nk_BT$ & $U^{sc}/Nk_BT$ \\ \hline
0.0 & $2.133\times10^{-5}$ &0& 0.20 &0  & & $0.0$              & $2.852 \pm 0.009$ \\
0.0 & $2.133\times10^{-5}$ &4& 0.20 &0  & & $-3.870 \pm 0.001$ & $3.056 \pm 0.005$ \\
0.4 & $2.133\times10^{-5}$ &0& 0.20 &0  & & $0.0$              & $2.363 \pm 0.010$ \\
0.4 & $2.133\times10^{-5}$ &4& 0.20 &0  & & $-8.152 \pm 0.001$ & $4.363 \pm 0.005$ \\ \hline
0.0 & $2.0\times10^{-4}$ &4& 0.20 &0 &  & $-6.411 \pm 0.003$ & $0.419 \pm 0.002$ \\ 
0.4 & $2.0\times10^{-4}$ &4& 0.20 &0 & & $-6.301 \pm 0.004$ & $0.388 \pm 0.002$ \\ \hline
0.0 & $2.133\times10^{-5}$ &40& 0.20 &0  & & $-82.924 \pm 0.023$ & $4.352 \pm 0.012$ \\
0.0 & $2.0\times10^{-4}$ &40& 0.20 &0 & & $-88.940 \pm 0.007$ & $2.242 \pm 0.004 $\\ \hline
0.0 & $2.0\times10^{-4}$ &4& 0.20 &2  & & $-6.871 \pm 0.003$ & $0.428 \pm 0.001$ \\ 
0.0 & $2.0\times10^{-4}$ &4& 0.20 &20  & & $-6.840 \pm 0.003$ & $0.424 \pm 0.001$ \\ 
\hline\hline
\end{tabular}}
\caption{Dimensionless energies in zero and uniform fields.}
\end{table}

\section{Structure deformation under external fields}
\label{deform}

In this Section we will discuss the effect of switching on a uniform field or field gradient on a pre-existing magnetic bijel. Recall that a uniform field creates only torques which cause dipoles to arrange along the direction of field; a field gradient creates both forces and torques. For this reason the two cases are quite different in their effects.


\subsection{Uniform field}

We first examine the effect of switching on a uniform field, on a pre-existing magnetic bijel. From an applications viewpoint, it is interesting to see whether a morphological transition, for instance from a bipercolating to a droplet or lamellar morphology, can be brought about by application of a field. To this end we consider a bicontinuous but asymmetric bijel with $\psi_0=0.3$ in a periodic box of $\Lambda^3=64^3$. (This is close to the percolation threshold; one might hope to see e.g. a loss of bicontinuity transverse to the applied field.) To maximise the chances of seeing such an effect, we choose magnetic colloids with extremely high dipolar interactions, $\lambda = 40$, retaining the usual volume fraction $\phi=0.2$. (We have also performed runs with lower $\lambda$; for $\lambda \le 8$ the effect of field on structure is very modest.) To see large effects we need also very high field strengths, with $\alpha_B = mB_0/k_BT$ in the range 50-1000. Although these $\lambda$ and $\alpha_B$ values are very high, they are not completely unfeasible: for instance, $\lambda = 40$ could be achieved using monodomain magnetite particles (magnetization density $4.8 \times 10^5$Am$^{-1}$) of 14.7nm radius, and for such particles $\alpha_B = 1000$ equates to $B_0\sim 0.63$T. This radius is close to the theoretical size limit for magnetite monodomains \cite{butler:1975}, beyond which particles become superparamagnetic rather than having permanent dipoles. As mentioned previously, however, in the presence of large fields as considered here, larger particles will anyway show saturating magnetization and while we have not simulated the superparamagnetic case, they should behave similarly to permanent dipoles with very large $\lambda$ values so long as the field remains in place.

\begin{figure}[tbph]
\centering
\vspace{.5in}
\subfigure[]{\label{fig6a}\includegraphics[width=0.4\textwidth]{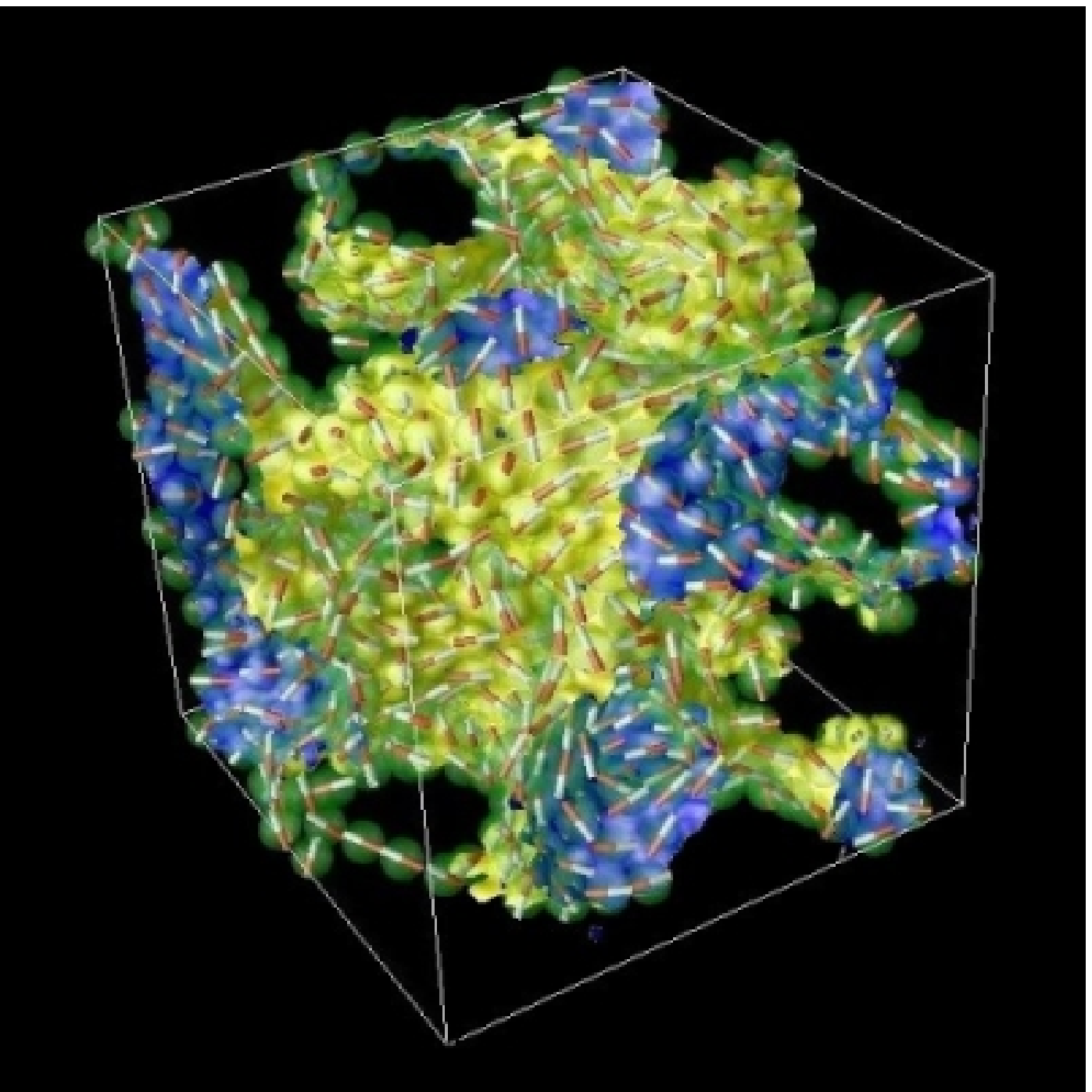}}
\subfigure[]{\label{fig6b}\includegraphics[width=0.4\textwidth]{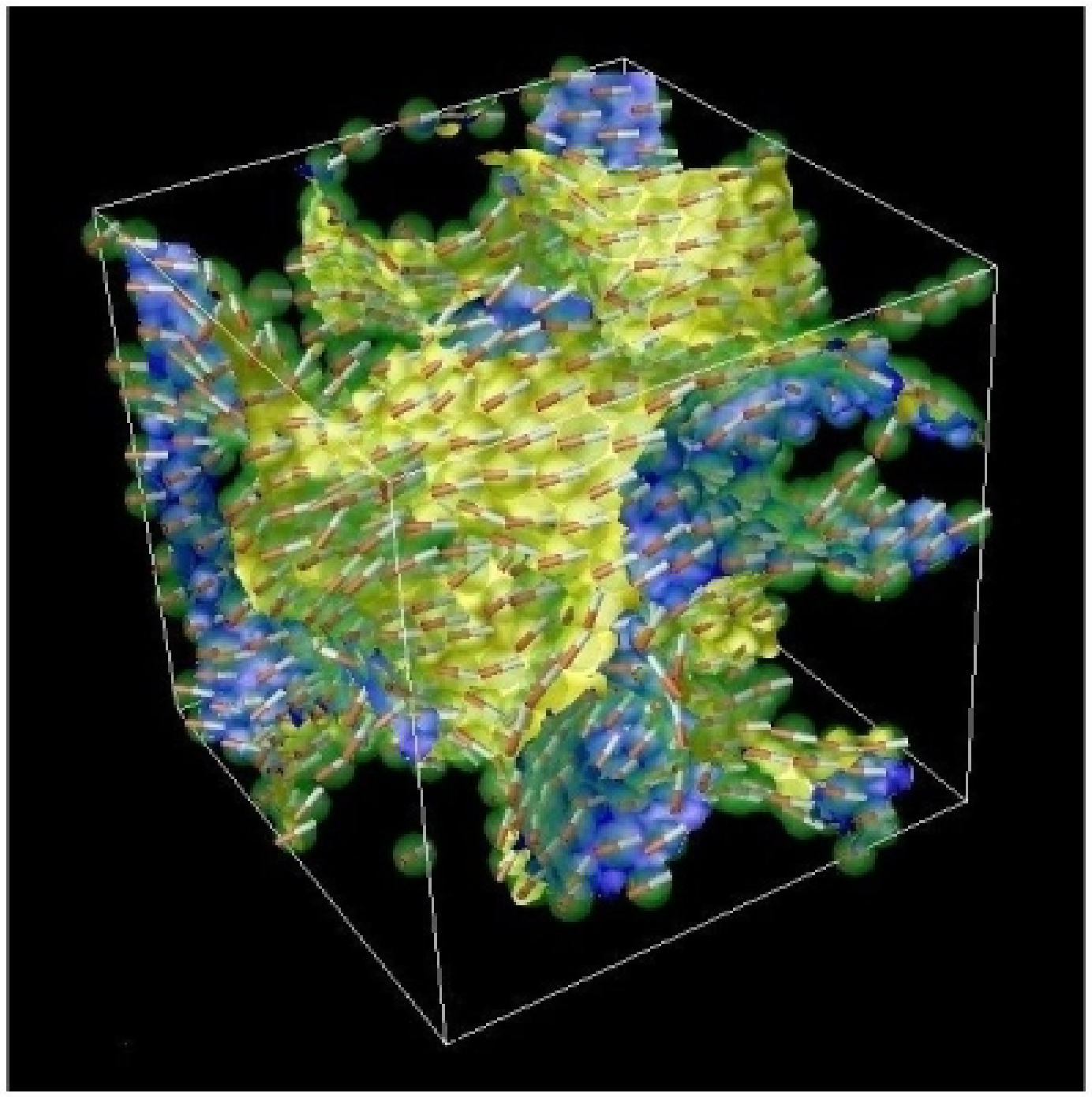}}
\subfigure[]{\label{fig6c}\includegraphics[width=0.4\textwidth]{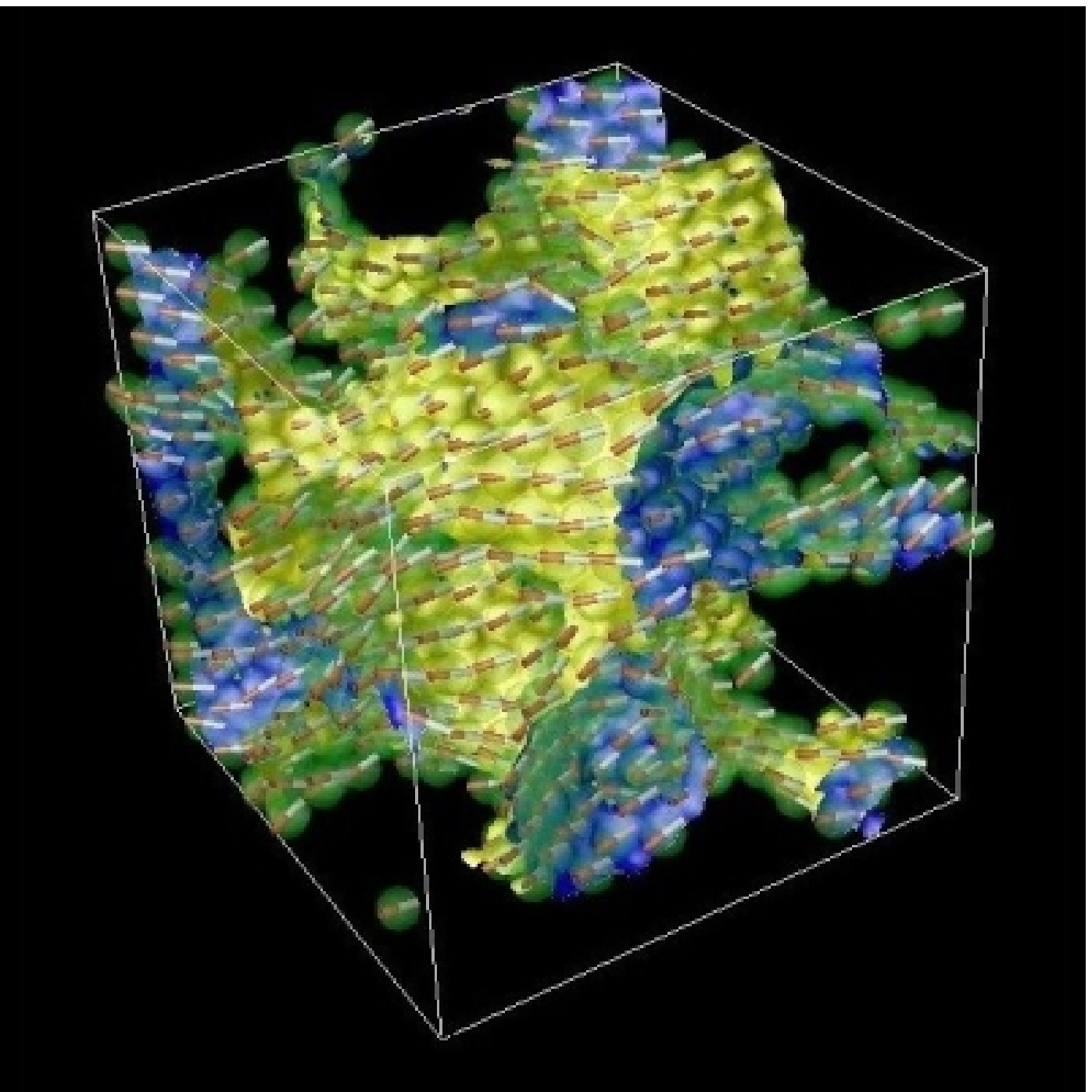}}
\subfigure[]{\label{fig6d}\includegraphics[width=0.4\textwidth]{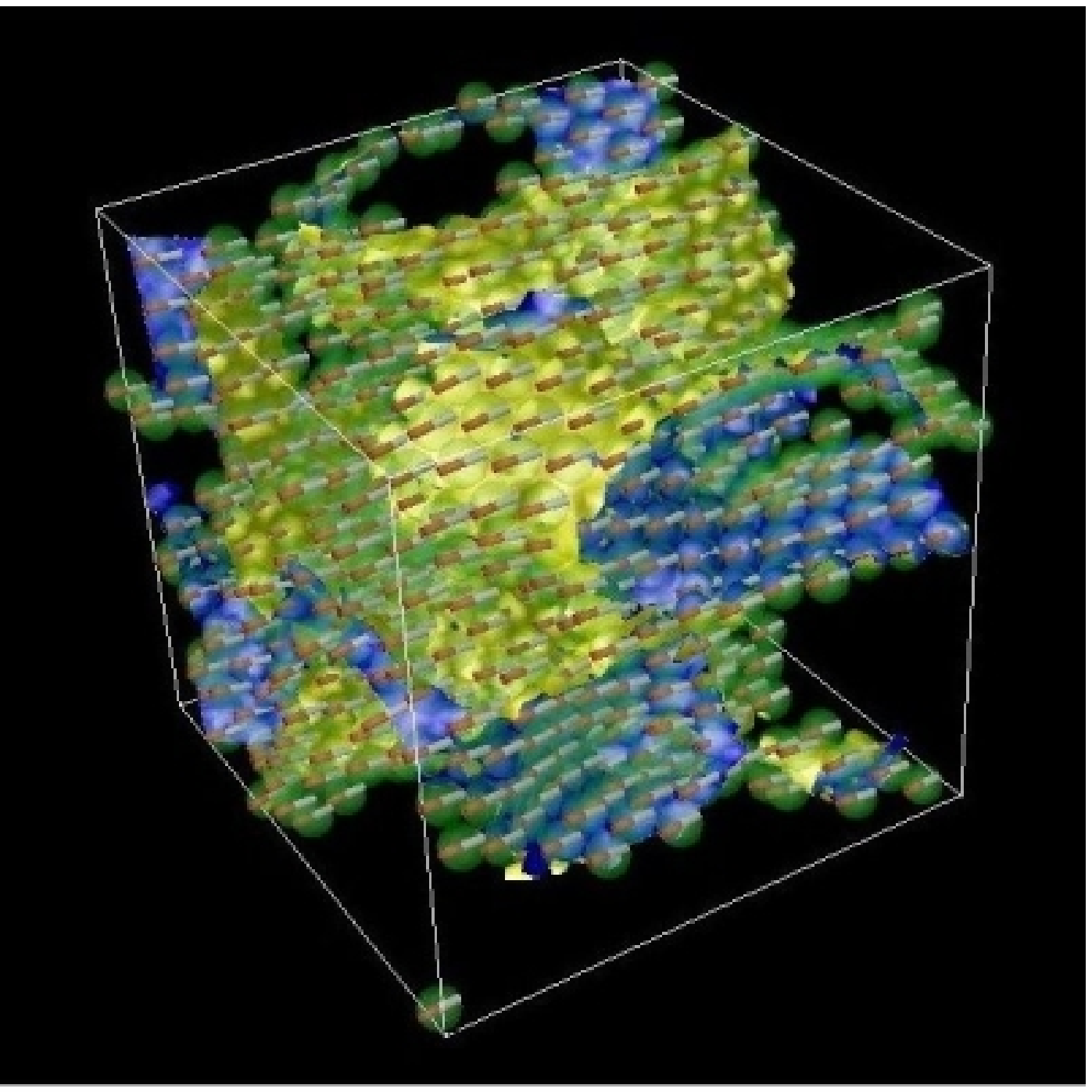}}
\subfigure[]{\label{fig6e}\includegraphics[width=0.3\textwidth]{fig6e.eps}}
\subfigure[]{\label{fig6f}\includegraphics[width=0.3\textwidth]{fig6f.eps}}
\subfigure[]{\label{fig6g}\includegraphics[width=0.3\textwidth]{fig6g.eps}}
\caption{Morphologies for $\psi_0=0.3$ with $\lambda=40$ at $k_BT = 2.133\times 10^{-5}$; system size $\Lambda^3=64^3$. (a) The initial stable configuration at $t=5\times10^5$ in zero field. The final configurations after switching on a field and then running for a further $5\times 10^5$ timesteps, with (b) $\alpha=50$, (c) $\alpha=100$, and (d) $\alpha=1000$. In panels (e)-(g) are shown the time evolution of characteristic length scales of fluid domains in $x$, $y$, and $z$ directions; the black dotted line is the initial run (a), the red solid line is for case (b), green dashed line (c) and blue dash-dot line is (d).}
\label{fig6}
\end{figure}

Figure \ref{fig6} shows snapshops of the final structures and graphs of the time evolution of the characteristic domain length-scales for $\alpha_B = 50,100,1000$. We have reverted here to the default temperature so that $\tilde \epsilon\equiv \epsilon/k_BT = 1230$. Thus the ratio of dipole-dipole to capillary energies is $\lambda/\tilde\epsilon = 0.0325$, and the ratio of field to capillary energies $\alpha_B/\tilde\epsilon$ lies in the range $0.04-0.81$. The field is applied along the $z$ direction which is here horizontal. The length scale plots show clear deformation of the structure; this takes some time to establish, but then appears to saturate. Even for the largest field we do not see depercolation of the bicontinuous structure in the final state, although this might possibly happen in a much larger sample, where the visible tendency to form horizontal sheets and/or tubes could eventually close off all fluid pathways transverse to the field.

\begin{figure}[tbph]
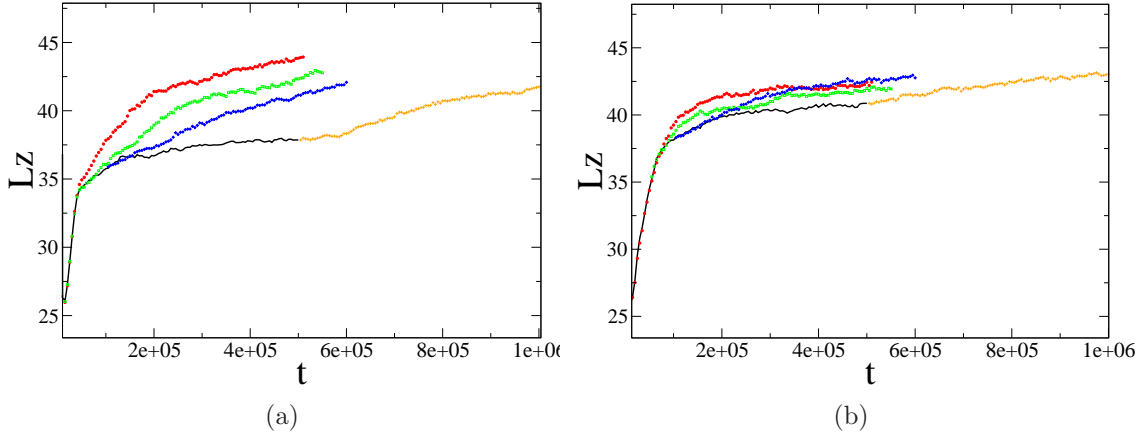

\centering
\vspace{.5in}
\subfigure[]{\label{fig7a}\includegraphics[width=0.45\textwidth]{fig7a.eps}}
\subfigure[]{\label{fig7b}\includegraphics[width=0.45\textwidth]{fig7b.eps}}
\vspace{.05in}
\caption{ The time evolution of the characteristic length scale along the direction of a uniform field of strength $\alpha_B = 50$, for various switch-on times $t_i$ of the field. Panel (a) is for $\lambda=40$ and (b) is for $\lambda=8$. The black curve is the growth rate of magnetic bijels without field; red, green blue, and orange are for $t_i=10^4$, $10^5$ and $5\times 10^5$ respectively.}
\label{fig7}
\end{figure}

The structures seen for $\alpha_B \ge 100$ are quite similar to the one shown in Fig.\ref{fig3}(c), for which the value of $\alpha_B/\tilde\epsilon = 0.15$ lies in the same range. This supports the idea that the field-induced effects on morphology are governed primarily by this ratio of field energy to capillary energy. Qualitatively, it does not seem crucial whether the field is applied during the initial formation of the bijel or switched on afterwards. To check whether this is quantitatively true, we show in Fig.\ref{fig7} the
evolution of the domain length $L_z$ along the field direction, for quenches at $\lambda = 40$ and $\lambda = 8$ in which an external field ($\alpha_B = 50$) was switched on at various times during the formation of the bijel. For the larger $\lambda$, at least, it seems clear that early switch-on does in fact promote significantly stronger anisotropy, presumably because the magnetic interactions (which result in chain formation) can in part template an anisotropic evolution of the interface before jamming is complete.

\subsection{Field gradient effects}

Our study of the effects of a field gradient is motivated by the experimental work referred to already \cite{fuller:2005/a}, in which a droplet emulsion stabilized by magnetic particles was controllably broken down by exposure to a field gradient. For small field gradients a slight elongation of the droplets was seen, but above a critical level the emulsion was completely destabilized leading to coalescence of droplets. We plan to address elsewhere \cite{elsewhere} the problem of modelling quantitatively this phenomenon for droplets, but here present data for the analagous situation involving bijels.

First, we generate the initial stable magnetic bijels in a closed-wall simulation box ($\Lambda^3=64^3)$ quenched at $\psi_0=0.3$. As usual we set $\phi=0.20$. The magnetic colloids again have a high dipolar coupling constant $\lambda=40$; we use our default temperature setting for which the dimensionless parameter describing the relative strength of dipolar and capillary forces is $\lambda/\tilde\epsilon = 0.0325$. To create a field gradient we introduce an external point magnetic dipole exterior to the box, which interacts with the colloidal dipoles; formally this involves addition of an extra particle in Eq.\ref{dipole} with a much larger $m$ value than any of the others. This is the simplest way to introduce a field gradient and automatically ensures $\nabla.{\bf B} = 0$ as required. So long as the additional dipole is placed far away from the box, the resulting field gradient is (almost) uniform.

\begin{figure}[tbph]
\centering
\vspace{.5in}
\includegraphics[width=0.95\textwidth]{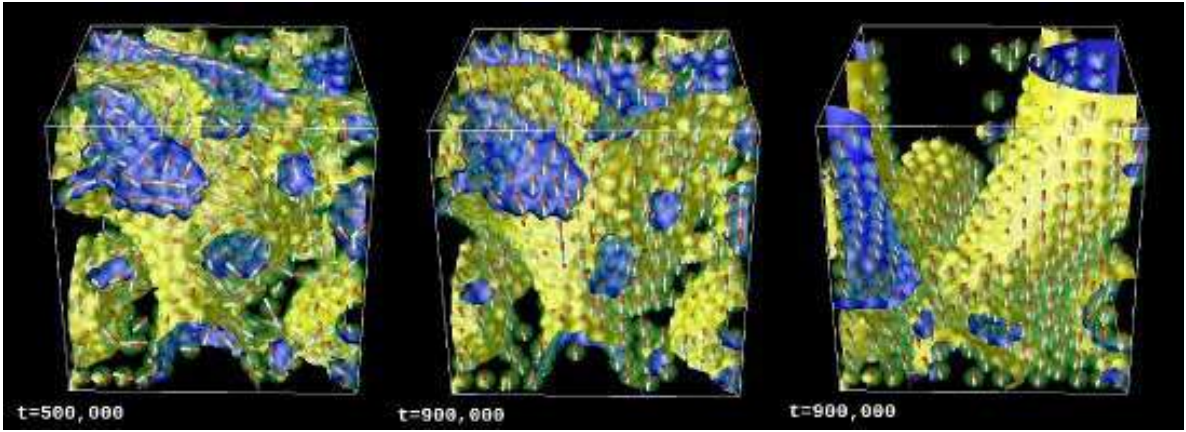}
\caption{Field gradient effects for $\lambda=40$ in a closed simulation box. From left to right, the initial snapshot at $t=5\times 10^5$; the  configuration after a further $4\times 10^5$ timesteps at dimensionless field gradient $C_1=0.01$ and $C_1=0.1$. The magnetic field is vertical.}
\label{fig8}
\end{figure}

The ratio of field-gradient to capillary forces for a single particle is given by a dimensionless number which we now define as $C_1 = m|\nabla B|/\pi \sigma a$. The dipole moments of the external magnet are $M=1.64\times10^6$ ($C_1=0.1$) and $1.64\times10^5$ ($C_1=0.01$) in our simulations, which correspond to $5.14\times10^{-11} \rm{Am^2}$ and $5.14\times10^{-12} \rm{Am^2}$ in the laboratory. The position of the external magnet is at (32, 32, -200) for a box size $\Lambda^3=64^3$.
We present in Fig.\ref{fig8} snapshots showing the result of switching on a field gradient for values $C_1 = 0.01$ and $0.1$. For the lower value the field gradient only weakly perturbs the bijel, which remains intact. For $C_1 = 0.1$, however, the magnetic particles are pulled off the fluid-fluid interface and accumulate at the bottom of the container in the region of maximum magnetic field.
The interface in the upper part of the container is deprived of particles, and coarsens until further growth becomes finite-size limited via interaction with the container walls.

This destabilization appears closely analagous to that reported by Melle et al \cite{fuller:2005/a} for droplets. To check whether it actually is similar, we need to compare the relevant dimensionless control parameter. We first assume this is the single particle force ratio $C_1$ as defined above. If as before we identify our simulations with $\lambda = 40$ as describing (hypothetical) monodomain magnetite particles of 17nm radius, then the required $C_1=0.1$ corresponds in the laboratory to a field gradient of roughly $4\times 10^7$ T/m. This is not feasible experimentally. In fact it is about seven orders of magnitude larger than the field gradient used in the experiments of \cite{fuller:2005/a}; however, these also used much larger particles (500 nm radius) of a superparamagnetic material. Assuming full saturation gives a value $C_1\simeq 10^{-5}$ at the point where the emulsion droplets are found to break in those experiments. 
 
Elsewhere we will however show that $C_1$ is not the relevant parameter to describe field-gradient-induced particle detachment in complex geometries \cite{elsewhere}. This is because the field gradient forces, which act like a body force on each particle, generally have a component tangential to the interface which is resisted by the interparticle repulsions through the contact network. This means that such forces can be transmitted over large distances and can in some circumstances act together in expelling a particle from the interface. Indeed, for the bijel in a fixed-wall container (Fig.\ref{fig8}) the net force on all particles present is resisted solely by the contacts between the bottom layer of particles and the basal wall. Accordingly, each of these particles feel a detachment force that is enhanced by a factor $N_{eff}$, the number of particles that it effectively supports against the field-gradient force, so that the relevant control parameter is not in fact $C_1$ but $C = C_1N_{eff}$. In an undistorted bijel geometry, we have $N_{eff} = N_T/N_B \simeq \Lambda/2a = 15$ where $N_T$ is the total number of particles in the simulation and $N_B$ is the number in contact with the base. Although the field-gradient induced breakup of a bijel requires unphysical field gradients in the simulations reported here,
$N_{eff}$ increases linearly with sample thickness. We cannot simulate very large samples, but if we did, we would expect the field gradient to fall accordingly towards the experimentally accessible range. Indeed, a laboratory bijel sample of $1$ cm thickness, made with particles of $500$nm radius, might have $N_{eff}\sim 10^4$ so that if breakdown occurs at $C = 0.1$ the threshold value of $C_1$ is about $10^{-5}$, close to that seen in the experiments of \cite{fuller:2005/a} on droplets.

\section{Discussion}
\label{conclude}

In this paper, we presented simulations of a binary fluid with magnetic colloids, undergoing a quench causing phase separation of the fluids. The neutral wettability of the colloids causes them to sequester at the fluid-fluid interface and then jam to form a bijel \cite{Stratford:2005/a,natmat}. For both a symmetric quench and an asymmetric quench, we found the basic morphologies remain broadly the same as for the systems with non-magnetic colloids. Dipole-dipole interactions cause head-to-tail alignment of the magnetic moments in the plane of the interface, whereas uniform external fields align all moments with the field direction. But, unless these interactions are simultaneously large, there is little change to either the interfacial particle packing or the fluid domain morphology that stems from it. These results show that, although dipole-dipole and dipole-field interactions are not negligible in determining the properties of magnetic bijels, for typical parameter values the dominant balance is that between capillary and short-range repulsive interactions, just as it is for the non-magnetic case. 

Such typical parameter values (such as $\lambda=4$ for the ratio of magnetic interactions to thermal energies) are those pertaining to magnetic nanoparticles that are already widely available for ferrofluid applications. In that context, extremely large $\lambda$ ($\sim 40$, say) may offer little advantage since such dipoles will aggregate in a quasi-irreversible fashion. In bijels, the magnetic interactions compete primarily with capillary forces, which are much larger than thermal ones, so the use of relatively large ferromagnetic particles may be appropriate. Moreover, since the structural changes reported here depend on the simultaneous presence of large dipolar interactions and large magnetic fields, in practice it may be possible to use superparamagnetic particles rather than the permanent dipoles studied in our simulations. If so, there is no longer any requirement to have a magnetic monodomain in each particle since the paramagnetic particles will anyway achieve saturating magnetization, creating in effect larger $\lambda$ than any achievable with monodomain (hence nanometric) particles. Indeed, the spectacular experiments of Melle et al. \cite{fuller:2005/a} on droplet emulsions use superparamagnetic particles of micron dimensions. Note that although it would be possible to perform simulations with $\lambda > 40$, there would be little chance for even local thermal equilibration of magnetic degrees of freedom \cite{kim:2009/a} and so we have not attempted this here. (Indeed for nanoparticles the effective run times of our simulations lie in the microsecond range \cite{Stratford:2005/a}: this is long enough for the bijel structure to undergo arrest but not to confirm explicitly that the structure is permanent -- for confirmation of which we appeal to experiment \cite{natmat}.)

When large fields and dipole-dipole interactions were both present, we observed significant anisotropy in the fluid domain morphology whose quantitative strength, but not qualitative features, depended on whether the external field was imposed from the start of the quench or switched on later. In this study we did not find a combination of dipolar and field parameters that would allow the macroscopic morphology to be switched dramatically, e.g. from a bicontinuous to a lamellar morphology. It remains possible that by a more complete exploration of parameter space such effects could be found. If so this would certainly be of interest for applications, allowing the permeability of the bijel to its two component fluids to be selectively controlled, in two different directions, by using a magnetic field.  

At modest dipolar interaction strengths we found strong frustration in the head-to-tail alignment and a slow relaxation of magnetic degrees of freedom. Even at the longest times simulated, significant frustration remained. Based on a careful examination of a single droplet arising within an asymmetric quench, we attributed this to an incompatibility between global head-to-tail alignment and defects in the particle packing (relative to a hexagonal 2D array) created by the curvature of the interfacial particle layer. These magnetically unfavorable defects can be locked in by the strong capillary forces which the dipolar interactions cannot overcome.  

At first sight, use of a magnetic field gradient (which can exert forces on particles) is less promising as a means of morphological control than that of a uniform field (which exerts only torques). Indeed the parameter $C_1 = m|\nabla B|/\pi \sigma a$, which controls the relative importance 
of field-gradient and capillary forces on a single particle, cannot be bigger than the parameter $mB/\pi\sigma a^2$ which governs the relative importance of magnetic and capillary torques in a uniform field. (To achieve equality, the field would have to fall from peak values to zero on the scale of one particle diameter.) However, unlike torques, the field-gradient forces can be transmitted by contact repulsions within the interfacial layer and can thereby gain a cumulative effect. Although in our simulations the resulting amplification factor is modest ($\sim 15$), for bijels we can see no limit to it in principle as the system thickness, in the direction of the field gradient, is increased. As discussed elsewhere \cite{elsewhere} related effects probably underly the experimentally field-gradient control of emulsion coalescence reported in \cite{fuller:2005/a}. It would certainly be interesting to conduct similar experiments on more complex interfacial assemblies stabilized by magnetic particles, and we hope that our predictions for these and other properties of magnetic bijels might stimulate such experiments in future.

\subsection*{Acknowledgments}
We acknowledge discussions with Philip Camp and Paul Clegg. This work was funded in part by EPSRC EP/030173 and EP/C536452. MEC holds a Royal Society Research Professorship.

\bibliographystyle{achemso}

\providecommand{\refin}[1]{\\ \textbf{Referenced in:} #1}

\end{document}